# Optical and magnetic properties of ZnCoO layers

M. Godlewski[1,2], M.I. Łukasiewicz[1], E. Guziewicz[1], V.Yu. Ivanov[1],

Ł. Owczarczyk[2], B.S. Witkowski[1]

[1]Institute of Physics, Polish Academy of Sciences, Al. Lotników 32/46,

02-668 Warsaw, Poland

[2]Dept. of Mathematics and Natural Sciences College of Science,

Cardinal S. Wyszyński University, Dewajtis 5, 01-815 Warsaw, Poland

**Abstract**

Optical and magneto-optical properties of ZnCoO films grown at low temperature by Atomic Layer Deposition are discussed. Strong wide band absorption, with onset at about 2.4 eV, is observed in ZnCoO in addition to Co-related intra-shell transitions. This absorption band is related to Co 2+ to 3+ photo-ionization transition. A strong photoluminescence (PL) quenching is observed, which we relate to Co recharging in ZnO lattice. Mechanisms of PL quenching are discussed.

# 1. Introduction

Wide band gap II-TM-VI and III-TM-V alloys (TM stands for a transition metal) are intensively studied for possible spintronic applications (see e.g. [1-4]). For some of them a room temperature ferromagnetism (RT FM) was theoretically predicted [2-4], which, if achieved, will result in several applications of these materials.

Whereas carrier mediated magnetic properties of GaMnAs are relatively well controlled, the present situation for ZnMnO and ZnCoO remains very unclear. RT FM was theoretically predicted in heavily p-type ZnMnO [2] and n-type ZnCoO alloys [3,4] and then observed experimentally. However, the origin of the observed magnetic ordering remains not clear (see e.g. [5] and references given there). It is more likely that RT FM of ZnCoO (also ZnMnO and GaFeN [6]) is due to inclusions of foreign phases and metal accumulations, rather than to "volume properties" of these two alloys (see e.g. [5] and references given there). We observed that samples with the uniform TM distribution show a paramagnetic response. Such samples were grown by us at a low temperature using Atomic Layer Deposition (ALD) [7,8]. They are used in the present study.

In the present work we investigate relation between optical and magnetic properties of ZnCoO layers grown at low temperature by the ALD. Some test measurements are also performed on bulk ZnO sample doped with Co, with a low Co concentration in the range of $10^{18}$ cm$^{-3}$. First we verify if we can realize conditions required for the FM ordering, as assumed in the theory. Then, we describe relations between optical and magnetic properties of ZnCoO (and also ZnMnO). We demonstrate that Mn and Co introduction to ZnO samples results in several new properties, not observed e.g. in ZnMnS [9]. In particular, the introduction of Mn and Co ions efficiently quenches visible emission of ZnO.

PL quenching is more efficient in ZnCoO, which was not surprising, since Co is known as a killer center in wide band materials, e.g. in ZnS [10]. Surprisingly, Mn ions also act as emission deactivators in ZnO, which is not observed in ZnS and many other II-VI compounds. This unexpected effect (in ZnS Mn is emission activator [9]) we utilize in practice to get information on samples uniformity. Efficient emission quenching in TM rich regions of a sample allows evaluation of uniformity of Co distribution in the samples studied. This information we obtain from maps of in-plane variations of the CL intensity, as already demonstrated for ZnMnO [11,12]. TM rich regions are observed as dark regions in the CL intensity maps.

In-depth and in-plane CL mapping performed by us evidence uniform Co distribution in the samples grown at low temperature by the ALD and nonuniform in the one grown at higher temperature, or annealed after the growth. Such experiments together with SIMS, XPS and magnetic investigations allowed us to find growth conditions to get ZnCoO layers with uniform Co distribution, as already discussed in the references [7,8,13] for ZnMnO layers. Uniform layers show a paramagnetic response at room temperature. A ferromagnetic one was detected in nonuniform films (grown at higher temperature) and was related to the presence of foreign phases and (dominant contribution) metal accumulations [5].

Despite the fact that a volume related ferromagnetism is not observed by us, the samples studied show some attractive magneto-optical properties, which we will relate to the observed non-uniformity in TM distribution and large carrier capture rates by Co (and Mn) ions in ZnO, as discussed below.

**2. Experimental**

Optical and magneto-optical investigations were performed on ZnCoO layers grown at relatively low temperature (LT, between 160 ºC and 200 ºC) by the Atomic Layer Deposition (ALD). Some test measurements were done on a bulk ZnO doped with Co, with Co concentration in the range of $10^{18}$ cm$^{-3}$. In the ALD process we used metalorganic zinc and manganese precursors - diethylzinc (DEZn) as zinc precursors and cobalt (II) acetyloacetonate - (Co(acac)$_2$) as cobalt precursor. As an oxygen precursor we used deionized water. We used 8:1 ratio of Zn-O to Co-O ALD pulses and low temperature of a growth (160 ºC), which we proved to be crucial to get uniform TM distribution in our alloys [5,7,8,13]. The so-obtained films were polycrystalline and contained about 1% of Co, uniformly distributed in plane and depth. Different substrates were used depending on experiment requirements – Si for magnetic investigations, (0001) sapphire or quartz for optical transmission investigations, glass for electrical measurements.

Most of investigations were performed on ZnCoO samples with relatively low Co fractions (about 1%). The larger Co fractions (about 7 – 8%) were observed by us for thin films grown with 2:1 ratio of Zn-O to Co-O ALD cycles at slightly increased temperature (about 200 ºC), for which we observed a FM response, as reported in [5]. FM response was related to Co metal accumulation at the Si/ZnCoO interface [5]. For thicker samples grown at a lower temperature and with a lower Co fraction we observed a paramagnetic response at room temperature and Co metal inclusions were absent or very rare [5]. EXAFS investigations (A.

Wolska, unpublished results) confirm that Co enters substitutional positions ($Co_{Zn}^{2+}$) in these ALD-grown ZnO samples. Some traces of metal inclusions were detected in XPS investigations of nonuniform samples (the one grown with 2:1 ratio of the ALD cycles and at 200 ºC) [5]. We also searched for inclusions of various oxides (Co oxides), as seen for us for studied in parallel ZnMnO films ($Mn_xO_y$), but their concentration (if they are present) must be very low. This is quite different situation to the one seen by us for ZnMnO. Co-OH formation was only detected in the XPS study.

Thickness of our samples was between 0.2 - 2 µm, which turned out to be crucial parameter for understanding magnetic properties of investigated samples. We observed a clear correlation between layer thickness and FM response for nonuniform ZnCoO layers [5]. Further details on growth conditions can be found in [5,7,8,13].

Investigations of RT transmission and photoluminescence (PL) spectra were performed using the Solar Spectrofluorimetr CM 2203 with two double monochromators, the Xe lamp as excitation source and the Hamamatsu photomultiplier (PMT). LT magneto-PL investigations were performed using a Spectromag 6000 split coil superconductive magnet of Oxford Instruments, He-Cd laser PLASMA model HCCL-15UM, a double monochromator equipped with Hamamatsu S7035 CCD camera and Hamamatsu photon counting system with R2531 PMT and FAST ComTec 7887 card. CL and SEM measurements were performed using Scanning Electron Microscope Hitachi SU-70 equipped with a GATAN MonoCL System.

RT Hall effect measurements were performed using the RH2035 PhysTech GmbH system equipped with a $B$=0.426 T permanent magnet. Electrical measurements were done in the van der Pauw geometry using e-beam evaporated Ti/Au as an Ohmic contact to ZnO.

## 3. Experimental results and their discussion

In absorption/transmission study of ZnCoO we observed the appearance of two types of Co-related absorption bands (see Figs. 1 and 2). In addition to a characteristic $Co^{2+}$ intra-shell transitions, being due to the $Co^{2+}$ intra - shell $^4A_2(^4F) \rightarrow {}^2A_1(^2G)$, $^4A_2(^4F) \rightarrow {}^2A_1(^4P)$ and $^4A_2(^4F) \rightarrow {}^2A_1(^2G)$ transitions, also reported by Koidl [14], Jin et al. [15], Dinia et al. [16], Liu et al. [17], and Ramachandren et al. [18], a broad absorption band is detected below the band-to-band transition. This is a similar situation as for ZnMnO, where we observed a broad band absorption below the onset of the band-to-band transitions, which was first reported by Fukumura et al. [19] and related to charge transfer transitions [19,20], smeared out the $Mn^{2+}$ intra-shell transitions [21], or to Mn 2+ to 3+ photoionization, as we discussed in details in our recent publication [22]. Transmission spectrum for the studied ALD-grown samples is shown in Fig. 2. The appearance of Co related charge transfer band is seen as in the bulk sample (see Fig. 1). For thin layers we also observed interference fringes superimposed on transmission spectra, as seen in Fig. 2. There the broad absorption band is so strong that it may result in an incorrect conclusion that the fundamental absorption (band-to-band) is reduced by Co introduction.

For ZnCoO a band overlapping the band-to-band transition was reported earlier by [23-29]. The origin of the latter absorption is still disputed. It was related to Co charge transfer transition [26,27,29], but also to excitation of surface plasmons resonances in nanometer size metallic Co inclusions [30]. The first interpretation was supported by the results of photoconductivity measurements [23-25].

The present study supports this identity of the absorption spectrum. We observe that PL excitation (PLE) within this absorption band results in recombination via $Co^{2+}$ intra-shell 3d states. This is not observed after the host excitation. Such property of the PLE spectrum is expected if the photogenerated free carriers recombine via a ladder of Co 2+ intra-shell states (electrons are retrapped by Co 3+ via excited Co 2+ states).

Origin of proposed charge transfer transitions was discussed by Liu et al. [26]. The authors proposed competition of two types of photoionization processes – Co 2+ to 3+ (at a lower energy) and 2+ to 1+ (close to the band-to-band transition), the latter overlapping with a process of photogeneration of Co localized excitons.

If the model proposed by Liu et al. is correct, films containing Cobalt should be resistive and high n-type conductivity could not be achieved in ZnCoO, i.e., the condition required for achieving a carrier related ferromagnetic coupling [3,4] could not be realized. $Co^{1+}$ level localized at about 0.28 eV [27] below ZnO conduction band edge should compensate shallow donors of ZnO. However, other authors reported possibility of achieving very high n-type conductivity in ZnCoO by Al codoping, which led to carrier related FM coupling [31]. Then the carrier mediated FM response was observed, but at low temperature in disagreement with the theoretical estimations. This indicates still a limited accuracy of first principle calculations.

In the present work we searched for correlations between Co fraction in ZnCoO layers and free electron concentration. Anti-correlation is expected for the 2+ to 1+ photoionization origin of the broad absorption band seen in Fig. 1.

Results of electrical investigations are quite confusing. We observe relatively high n-type conductivity for some of the layers, as well as a high resistivity for other ZnCoO films. Reasons for this observation are still not clear. For example, keeping the same growth temperature and number of the ALD cycles, but increasing length of Co precursor doses, we can change the sample resistivity from a low one to a very high in samples grown using longer times of Co doses. These samples were of the same thickness, contained similar Co fraction but had very different electrical properties (from free electron concentration of a few times $10^{18}$ cm$^{-3}$ to resistive samples).

One of possible explanations is formation of Co-H-Co bonds in ZnCoO, as theoretically predicted [32]. This reduces concentration of shallow donors and thus may result in resistive samples. If such explanation is correct, the observed reduction of n-type conductivity in ZnCoO films is not related to Co 2+ to 1+ recharging, but to formation of Co-H (compensation of H shallow donors) complexes. However, we do not observed any correlation between Co and H fractions in the samples, and XPS study indicate only formation of Co-OH bonds.

Liu et al. [26] assumed two photoionization transitions in ZnCoO - 2+ → 1+ and 2+ → 3+. Possible charge states of Co in ZnO should be detected with electron spin resonance (ESR). If the model proposed by Liu et al [26] is correct, in n-type samples Co should be in 1+ charge state, in resistive samples in 2+, and in p-type in 3+. Co$^{2+}$ ESR signal was reported in several cases (see e.g. [33-35]). ESR signal of Co$^{1+}$ (3d$^8$) has been observed in MgO, CaO and hexagonal CdS [36] but not in ZnO. Also Co 2+ intra-shell absorption should not be seen in n-type samples, which is not the case, see Figs. 1 and 2. There Co should be in 1+ charge state. Thus, it is unlikely that the broad band absorption with the onset at about 2.4 eV is due to 2+ to 1+ Co transition. Also attribution of this band to excitation of surface plasmons resonances is unlikely. Co-related below band gap absorption was also seen for uniform samples showing a paramagnetic response, with no Co metal inclusions detected by us in the XPS investigations. We conclude that the 2+→3+ Co recharging takes place in ZnO. This interpretation is supported by the theory, since the 3+ charge state of Co in ZnO was predicted theoretically [37].

Mn, Co recharging is crucial to explain the observed by us PL deactivation in ZnCoO and also in ZnMnO [11,12,28]. CL intensity is reduced in samples with increasing Co (Mn) fractions, which led us to conclusion that both Mn and Co act as killer centers in ZnO. To verify this statement we carefully check if the CL intensity variations are not related to samples microstructure only. To achieve a uniform TM distribution our samples were grown at a LT on costs of their structural properties. Samples have a polycrystalline structure when grown at low temperature. We applied often used method to improve sample quality and enhance light emission. A post-growth annealing (we used rapid thermal annealing in nitrogen gas) was applied to get a bright light emission, in particular in the band edge spectral region.

In Fig. 3 we show the results of CL investigations of ZnCoO sample as-grown and post-growth annealed up to 800 $^{o}$C. Annealing in fact helps to get a brighter light emission, but if we exceeded 600 $^{o}$C annealing temperature a new effect is observed – annealing results in Co redistribution, and in the appearance of dark regions in the CL maps (see Fig. 3 for ZnCoO samples showing uniform Co distribution before annealing). CL emission is deactivated in the regions of an increased Co concentration, as we also observed for ZnMnO [11,12,28]. We independently verified nonuniform TM (Mn/Co) distribution by comparing the results of the CL depth profiling investigations and the SIMS investigations [38.].

If the extra absorption band shown in Figs. 1 and 2 is due to Co ionization, we can explain many of the puzzling ZnMnO and ZnCoO magneto-optical properties, which we previously described for Mn ions in ZnO lattice [11,39]. Firstly, emission predominantly comes from the regions of a reduced TM concentration, which explains often observed lack of expected strong magneto-optical effects [11,38,39] expected for diluted magnetic semiconductors (DMSs). For example, we do not observe enhancement of the Zeeman splitting of excitonic transitions. Secondly, PL shows a circular polarization (see Fig. 4), which is another interesting magneto-optical effect, related by us below to the mechanisms responsible for the PL quenching by TM ions. This effect is much stronger in the case of ZnMnO [38] (see Fig. 4) than for ZnCoO. For ZnCoO circular polarization of excitonic PL is about 15 %, whereas for ZnMnO PL polarization exceeded 35% at 6 T. The difference we relate to TM concentration in the studied samples. In the case of ZnMnO PL/CL was quenched in samples with Mn fraction above 5%. Thus, we used samples with Mn fractions up to 5 %. For ZnCoO magneto-optical study was possible only for the samples with a very low Co fraction (in most cases below 1%), since for ZnCoO the effect of emission quenching is by far more efficient. Thus, PL investigations were possible for the samples with Co concentration below 0.5 – 1 %.

Mechanisms of visible PL deactivation in wide band gap II-VI semiconductors were discussed by us for ZnFeS [40,41] and ZnFeSe [42]. Quenching of a visible PL was related to the competition of three mechanisms (see [43] and references given in): 1) to the so-called bypassing effect [10] - efficient carrier recombination via a mid band gap TM-related level; 2) to the Auger-type energy transfer from excitons and recombining donor-acceptor pairs (DAPs) to TM ions (energy is used for a TM ionization), and; 3) to formation of complex centers of TM ions with common PL activators. Two of the former processes (bypassing and Auger effect) are most likely responsible for the PL deactivation in the present case. However, if Co-H-Co bonds are formed, also the third mechanism may be important in ZnCoO.

The presence of the two Co (and Mn) charge states in ZnO may result in a new interesting magneto-optical property of the samples studied. Even in samples with relatively low Co (Mn) fractions (low TM concentration is required to get efficient light emission) we observe relatively strong circular PL polarization, as we already reported for ZnMnO. A very strong PL circular polarization of a similar origin was seen by us also for Cr doped ZnSe and ZnTe [44] and was related to spin selective carrier retrapping via TM ions. High efficiency of the bypassing process means that capture cross sections for free carrier trapping by deep Co (also Mn) related levels are so high, that these carriers very efficiently recombine via TM levels. In a magnetic field carrier trapping becomes spin dependent. Carriers with a given spin polarization recombine via Co (Mn) deep level, whereas remaining carriers (the one with an opposite spin orientation) form excitons or are trapped by donors and acceptors. This results in a circular PL polarization, even though other magnetic effects expected for DMS samples are not observed.

**Conclusions**

Concluding, optical investigations of ZnMnO and ZnCoO favor the model of mix valence of Co (and Mn) ions in ZnO. The proposed 2+ to 3+ origin of the photoionization transition means that we can achieve high n-type conductivity of ZnCoO, but not p-type. In accordance to theoretical predictions electrons mediated FM is still possible in ZnCoO, but it is unlikely that we will rich RT FM. Experimental results given in the reference [31] indicate overestimation of carrier mediated interactions in ZnCoO. On other hand, high efficiency of the bypassing process results in a possibility of achieving circular polarization of light emission in the magneto-optical investigations. The effect, formally expected in samples showing magnetic ordering, is observed as the consequence of a large efficiency of the bypassing process.

**Acknowledgments**

This work was partly supported by FunDMS ERC Advanced Grant Research and by the European Union within European Regional Development Fund, through Grant Innovative Economy (POIG.01.01.02-00-008/08).

**Figure captions:**

**Figure 1:** Absorption spectrum of the reference bulk ZnO:Co sample measured at 2 K temperature.

**Figure 2:** Room temperature transmission spectrum for the ALD-grown ZnCoO layer.

**Figure 3:** Correlation between sample microstructure (from the SEM image (left) and in-plane variations of PL intensity (from the cathodoluminescence investigations, right). Data were taken for ZnCoO layer after a post-growth annealing at 800 $^{o}$C.

**Figure 4:** Circular polarization of the band edge emission of ZnMnO and ZnCoO (shown in the inset) measured at 2K and magnetic field up to 6 T.

**Figure 1:** Absorption spectrum of the reference bulk ZnO:Co sample measured at 2 K temperature.

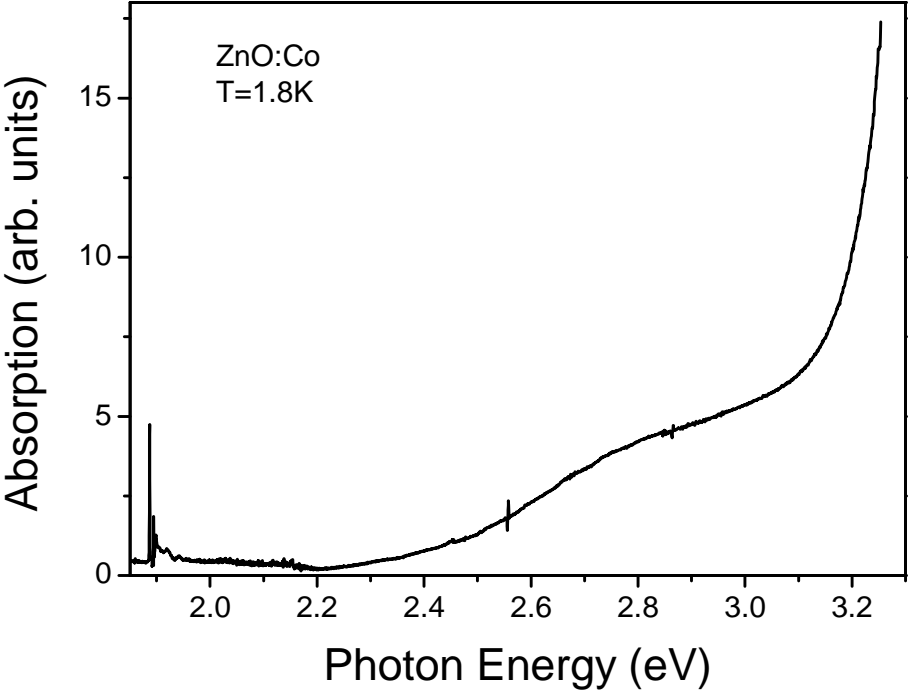

**Figure 2:** Room temperature transmission spectrum for the ALD-grown ZnCoO layer.

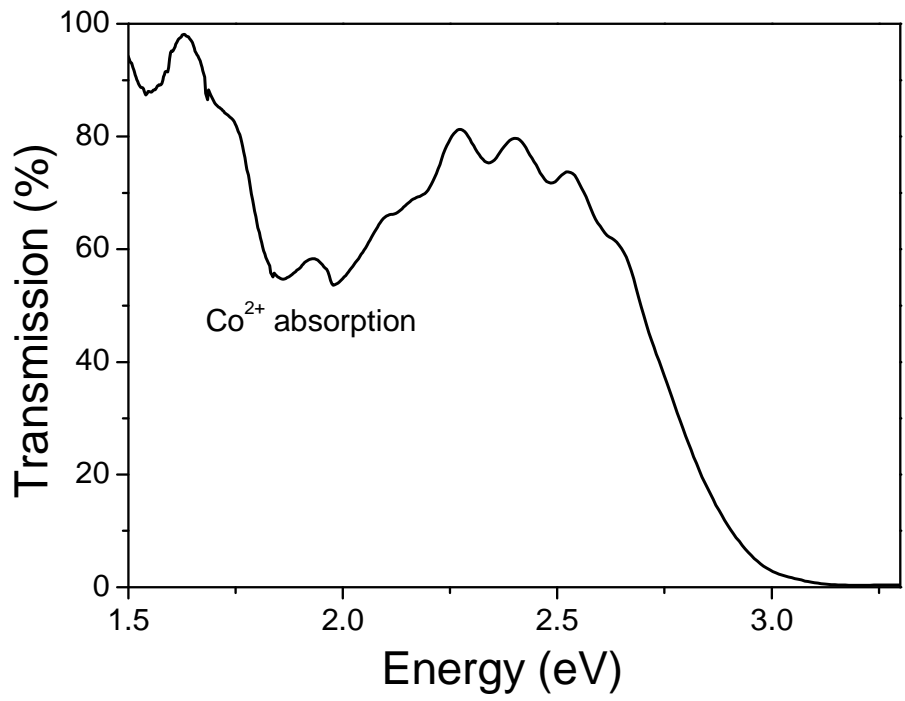

**Figure 3:** Correlation between sample microstructure (from the SEM image (left)) and in-plane variations of PL intensity (from the cathodoluminescence investigations, right). Data were taken for ZnCoO layer after a post-growth annealing at 800 °C.

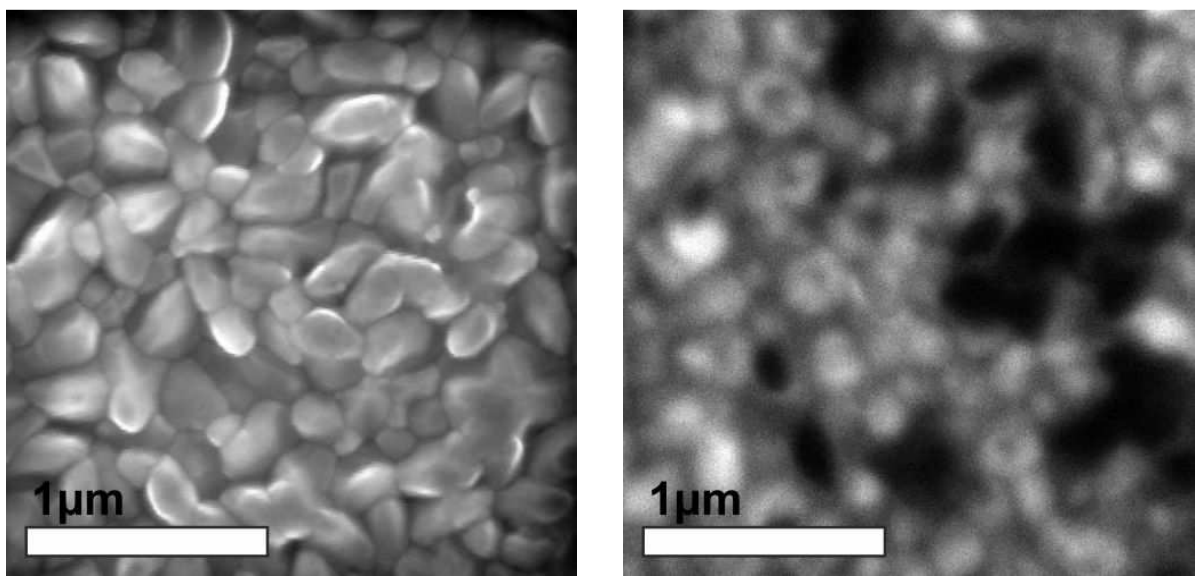

**Figure 4:** Circular polarization of the band edge emission of ZnMnO and ZnCoO (shown in the inset) measured at 2K and magnetic field up to 6 T.

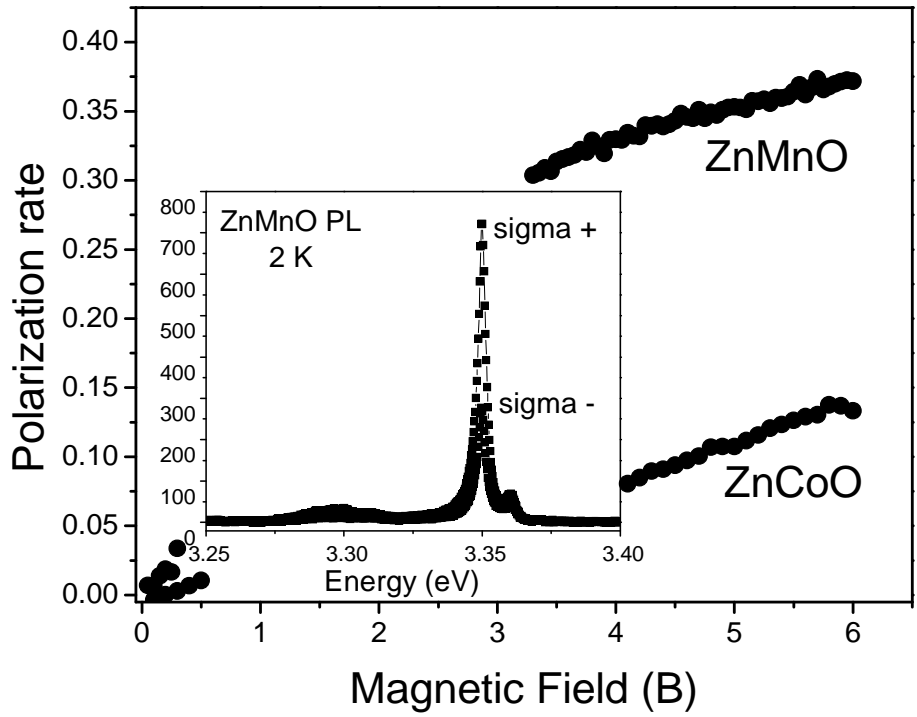